\documentclass[onecolumn,draftclsnofoot,12pt]{IEEEtran}
\usepackage{cite}
\usepackage{xr}
\usepackage{subcaption}
\ifCLASSINFOpdf
\usepackage[pdftex]{graphicx}
\DeclareGraphicsExtensions{.pdf}
\else
\fi

\usepackage{mathtools}
\usepackage{graphicx}
\usepackage{amsthm}
\usepackage{cite}
\usepackage{color}
\usepackage{bbm}
\usepackage{amsmath,amssymb,amsfonts}

\newcommand{\x}{\pmb{x}}

\newcommand{\ptwo}[2]{\mathrm{P}_{#1}(#2)}
\newcommand{\pone}[2]{\bar{\mathrm{P}}_{#1}(#2)}
\newcommand{\dd}{\mathrm{d}}
\newcommand{\ISI}{I}
\newcommand{\microm}{\ \mu\text{m}}
\newcommand{\xaxis}{$\mathsf{x}$}
\newcommand{\yaxis}{$\mathsf{y}$}
\newcommand{\pred}[2]{\mathrm{s}_{#1}(#2)}
\newcommand{\var}[2]{\mathrm{Var}\left(#1\right)}
\newcommand{\lptwo}[2]{\mathcal{P}_{}(#2)}
\newcommand{\lpone}[2]{\overline{\mathcal{P}}_{}(#2)}
\renewcommand{\ptwo}[2]{\mathrm{p}_{}(#2)}
\newcommand{\ptotal}[2]{\mathrm{p}_{\mathrm{T}}(#2)}
\renewcommand{\pone}[2]{\overline{\mathrm{p}}_{}(#2)}
\newcommand{\ie}{{\em i.e.}~}
\newcommand{\pdetect}[2]{\mathrm{P}_{\mathrm{d},#1}(\eta_{#1},#2)}
\newcommand{\pfalse}[2]{\mathrm{P}_{\mathrm{f},#1}(\eta_{#1},#2)}

\newcommand{\auc}[2]{\mathrm{A}_{#1}[#2]}

\newcommand{\ts}{T_{\mathrm{S}}}
\newcommand{\norm}[1]{\left\lVert #1\right \rVert}

\newcommand{\erfc}{\mathrm{erfc}}
\newcommand{\erf}{\mathrm{erf}}
\newcommand{\expect}[2]{\mathbb{E}_{#1}{\left[#2\right]}}
\newcommand{\pro}{\mathbb{P}}
\newcommand{\rthree}{\mathbb{R}^3}

\newtheorem{coro}{Corollary}

\usepackage{verbatim}
\begin{document}
	
	\title{\LARGE{3-D Diffusive Molecular Communication with Two Fully-Absorbing Receivers: Hitting Probability and Performance Analysis}}
	\author{Nithin V. Sabu,  Neeraj Varshney 
		and Abhishek K. Gupta
		\thanks{ N. V. Sabu and A. K. Gupta are with Indian Institute of Technology Kanpur, Kanpur UP 208016, India (Email:{ \{nithinvs,gkrabhi\}@iitk.ac.in}). 
				N. Varshney is with the Wireless Networks Division, National Institute of Standards and Technology, Gaithersburg, MD 20899 USA (Email: {neerajv@ieee.org}).
		}
}
	
\maketitle

\begin{abstract}
Exact analytical channel models for molecular communication via diffusion (MCvD) systems involving multiple fully absorbing receivers (FARs) in a three-dimensional (3-D) medium are hard to obtain due to the mathematical intractability of corresponding diffusion equations. 
Therefore, this work considers an MCvD system with two spherical FARs in a 3-D diffusion-limited medium and develop several insights using an approximate analytical expression for the hitting probability of information molecule (IM). Further, based on the hitting probability, a novel approximate closed-form analytical expression for the area under the receiver operating characteristic curve (AUC) is derived to analyze the detection performance at each FAR in the presence of other FAR. Finally, simulation results are presented to validate the analytical results using the particle-based and Monte-Carlo simulations and to yield important insights into the MCvD system performance with two FARs.

\end{abstract}	
\section{Introduction}
 In recent times, the Internet of Bio-Nano Things (IoBNT) is gaining
significant prominence towards addressing challenging problems
in biomedical scenarios, where multiple transmitters and
receivers have to work together to perform complex tasks,
including sensing and actuation \cite{akyildiz2015internet}.  In this context, molecular communication via diffusion (MCvD) has gained significant research attention to realize communication between bio-nano-machines within the IoBNT. In an MCvD system, information molecules (IMs) emitted from the transmitter propagates to the receiver via Brownian motion\cite{Jamali2019}.\par
\textit{Related Work}: In the context of multiple devices communicating using MCvD,  \cite{Kwack2018} considers an 
MCvD system with two FARs in a 3D medium and obtains the hitting probability of an IM at each fully-absorbing receiver (FAR), or equivalently the average fraction of IMs absorbed by each FAR. Note that in contrast to passive receiver where the receiver does not affect the propagation of IMs, a FAR immediately absorbs the IMs once they hit its  surface\cite{Jamali2019}.  
The channel model derived in \cite{Kwack2018}  consists of two unknowns, which have to be computed numerically. To the best of our knowledge, the exact expressions of the hitting probability for systems with multiple FARs are not available in the literature. Due to the lack of analytical channel model for an MCvD system with multiple FARs in 3D medium, most of the works \cite{Lu2016,Bao2019,Koo2016}
relied on simulation-based channel models to analyze the system performance. 
 In the past, \cite{Berezhkovskii1990} studied the 3D kinetics of a  Brownian particle in the presence of two spherical traps and presented an approximate expression for the death probability of this particle from any of these two traps. This analytical framework can be applied to derive channel models for MCvD systems with two FARs and study its performance, which is the prime focus of this letter.
The derived channel model has applications in a variety of interesting scenarios, including (1) interference and performance analysis of an MCvD system with two FARs (since transmitters as well as absorbing receivers can cause interference in the communication) and (2) MIMO systems with two FARs that act as receiving units to a common receiver.

\textit{Contributions}: In this letter, we consider an MCvD system with one transmitter and two spherical FARs in a diffusion-limited 3D medium and develop a channel model based on the analysis in  \cite{Berezhkovskii1990}. Using the proposed channel model, we further develop an analytical framework to study such systems.  We then validate the presented analysis via  particle-based simulations. We provide several design insights related to the mutual influence of FARs and  their mutual distance's impact on the hitting probability of an IM on each FAR. 
We also derive the area under the receiver operating characteristic curve (AUC) for each individual FAR, which serves as a quantitative measure for the FAR's capability of correct decision and study the impact of the presence of another FAR on it. We also provide a novel approximate closed-form analytical expression for  AUC.
\section{System Model}
 \begin{figure}
	\centering
	\includegraphics[width=0.7\linewidth]{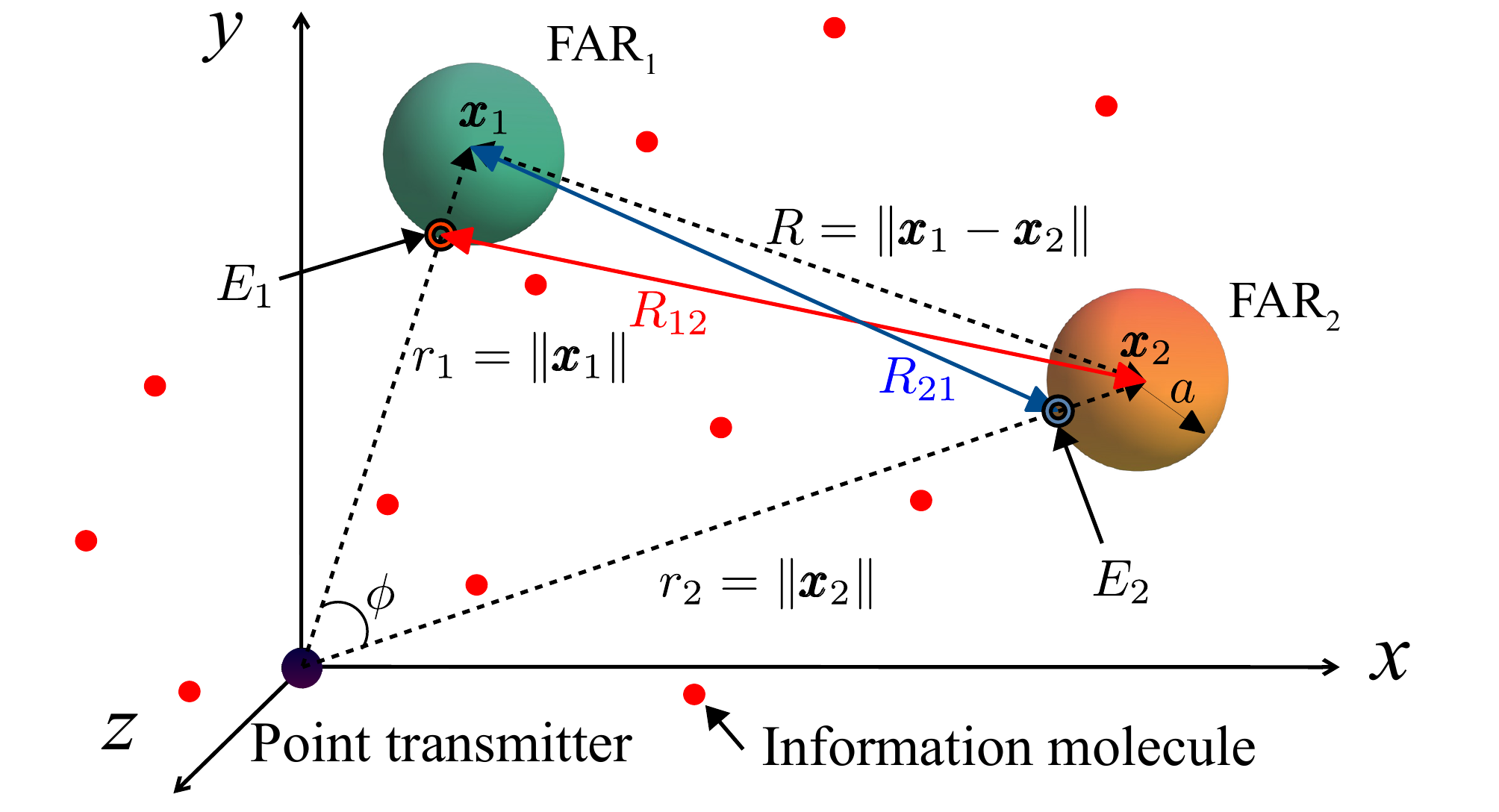}
	\vspace{-0.1in}
	\caption{A 3D MCvD system with a point transmitter at the origin and two FARs located at $ \x_1 $ and $ \x_2 $.  $E_i$ denotes the  closest point of  $i$th FAR from the transmitter.}
	\label{fig:sm}
	\vspace{-0.2in}
\end{figure}
 In this letter, we consider an MCvD system with a point
 transmitter and two  spherical FARs at different locations in a 3D
 medium as shown in Fig. \ref{fig:sm}.
 Let the  transmitter be located at the origin and the two FARs of radius $a$ at positions $\x_1$ and $\x_2$ respectively in $\rthree$ space.
The time  is divided into  time-slots of duration $\ts$, \ie $l$th time-slot denotes the time period $[(l\ts,(l+1)\ts]$ with $l\ge 0$. 
At the beginning of each time-slot, the  transmitter transmits its binary information using the on-off keying  modulation, \ie the transmitter emits $N$ IMs for bit $1$ and does not emit any IMs for bit $0$.
 IMs have a diffusion coefficient $D$ with respect to the propagation medium. We assume $D$ to be constant over the space and time. In any time-slot $l$, the transmit bit $b[l]$ is an independent Bernoulli random variable taking value 1 with probability $q_1$, and  $0$ with $q_0=1-q_1$. 
The transmitter and both FARs are assumed to be synchronized in time, which is a common assumption in the past literature \cite{Koo2016}.

The probability that an IM reaches (and gets absorbed by) the $i$th FAR located at $\x_i$ within time $t$ in the presence of the other FAR located at $\x_j$, is denoted by $\ptwo{i}{t,a,r_i,r_j}$. Here $r_i=\norm{\x_i}$ and $r_j=\norm{\x_j}$. This  {\em hitting probability}  equals the average fraction of IMs absorbed until time $t$ by the $i$th FAR. 
Let us now focus on a particular time-slot $l$. Now, the probability that an IM emitted at the $k$th slot reaches in the $ l $th slot at $i$th FAR  can be written in terms of $\ptwo{i}{t,a,r_i,r_j}$ as
\begin{align}
\!\!h_i[l-k]=&\ptwo{i}{(l{-}k{+}1)\ts,a,r_i,r_j}\!\!-\!\ptwo{i}{(l{-}k)\ts,a,r_i,r_j},\!\!   \label{FHTP}
\end{align}
\noindent$ \forall l\ge k $. In particular, $h_i[0]$ denotes the probability that IM transmitted in $l$th time-slot arrives in the same time-slot.  
We assume that there are no potential collisions between the IMs during their propagation in the medium \cite{Jamali2019}, and hence, the motion of an IM is independent of the motion of other IMs. 
Thus,  $S_i[l]$ denoting the number of IMs reaching the FAR$_i$ in the  $l$th time-slot corresponding to bit $b[l]$ is Binomial distributed with parameters $Nb[l]$ and $h_i[0]$, \ie $S_i[l]\sim\mathcal{B}( Nb[l],h_i[0])$.
Note that if $n$ is large, $\mathcal{B}(n,q)$ can be approximated as Gaussian distribution $\mathcal{N}(\mu,\sigma^2)$ with mean $\mu=nq$ and variance $\sigma^2=nq(1-q)$ \cite{Jamali2019}.  
Hence, we can model 
$S_i[l]\sim\mathcal{N}(Nb[l]h_i[0],Nb[l]h_i[0](1-h_i[0]))$. %
Similarly, $I_i[k]$ denoting the number of IMs received at the FAR$_i$ in $l$th time-slot corresponding to the transmission
in $k$th slot ($k<l$) can be modeled as 
\begin{center}
$ I_i[k]\sim\mathcal{N}\left(Nb[k]h_i[l-k],Nb[k]h_i[l-k](1-h_i[l-k])\right)$. 
\end{center} 
Note that, $ I_i[k] $ corresponds to inter symbol interference (ISI) arising due to the transmission from previous $k$th time-slot. Now, the total number of IMs corresponding to all previous transmissions 
 is $\ISI_i[l]=\sum_{k=1}^{l-1}I_i[k]$.  Let  $N_i[l]$ denote the number of molecules received from  unintended sources with $N_i[l]\sim \mathcal{N}(\mu_\mathrm{n},\sigma_{\mathrm{n}}^2)$ \cite{Meng2012}. 
Now, $Y_i[l]$ denoting the total number of molecules arriving at  FAR$_i$ in the  $l$th time-slot, is given as
\begin{align}
	Y_i[l] = S_i[l]+\ISI_i[l]+N_i[l].   \label{RecMol}
\end{align}
During detection, FAR$_i$   decodes $\widehat{b}_i[l]=1$ when $Y_i[l]\geq \eta_i$ and $\widehat{b}_i[l]=0$ otherwise, where  $\eta_i$ is the  decision threshold. 
Before proceeding further, we will calculate  the mean $\mu_{b[l]}[i;l]$ and  variance $ \sigma_{b[l]}^2[i;l] $  of the random variable $ Y_i[l] $ for $  b[l]\in\{0,1\} $, which would be useful when analyzing the performance of the receiver in Section \ref{sec:auc}.  These  can be derived as
	\begin{align*}
	\mu_0[i;l] 		= &	Nq_1\sum_{k=1}^{l-1}h_i[l{-}k] + \mu_\mathrm{n},\\
	\sigma_0^2[i;l] 	= &  	Nq_1\sum_{k=1}^{l-1} 
								\Big[ h_i[l{-}k](1{-}h_i[l{-}k])+
								Nq_0h_i[l{-}k]^2\Big]{+}\sigma_\mathrm{n}^2,\\
	\mu_1[i;l] 		= &	Nh_i[0]+\mu_0[i;l], \\
	\sigma_1^2[i;l] 	= &	Nh_i[0](1{-}h_i[0])+\sigma_0^2[i;l]. 
	\end{align*} 	
In contrast to Genie-aided approach where the means and the variances are obtained in terms of  previous bits \cite{genie2}, we assume  the previous bits to be random and take average over them for the above calculations 
\cite{avg2}.
In the next section, we will discuss  the hitting probability  $\ptwo{i}{t,a,r_i,r_j}$ for each of the FARs and  their influence on each other.
\section{Mutual Influence of the Two FARs}
The exact analytical expression for $\ptwo{i}{t,a,r_i,r_j}$ in a 3D medium is not available in the existing literature due to its intractability.
However,  using the analytical framework given in \cite{Berezhkovskii1990}, an approximate value for it can be obtained as,
\begin{align}
\ptwo{i}{t,a,r_i,r_j}=&\sum_{n=0}^{\infty}\frac{a^{2n}}{R_{ij}^nR_{ji}^n}\left[\frac{a}{r_i}
\erfc\left(\frac{r_i-a+n\left(R_{ji}-a\right)+ n\left(R_{ij}-a\right)}{\sqrt{4Dt}}\right)\right.\nonumber\\
&\left.-\frac{a^2}{r_jR_{ji}}
\erfc\left(\frac{ r_j-a+(n+1)\left(R_{ji}-a\right)+ n\left(R_{ij}-a\right)}{\sqrt{4Dt}}\right)\right],\label{e4}
\end{align}

\noindent where $R_{ji}$ is the distance between the center of $i$th FAR and the closest point of $j$th FAR from the origin ($ \text{E}_j $, see Fig. \ref{fig:sm}). If  $ \phi $ is the angle between vectors $ \x_1$ and $ \x_2 $, 
 then 
$R_{ji}=\sqrt{(r_j-a)^2+r_i^2-2(r_j-a)\cdot r_i\cos(\phi)}$. 
The approximation is good under the assumptions- (i) the distance between the transmitter and each FAR is significantly larger than $a$, 
 \ie $r_1\gg a$ and $r_2\gg a$, and (ii) the distance between  FAR$_1$ and the FAR$_2$ is significantly larger than $ a $, \ie  $ R=\norm{\x_i-\x_j}\gg a$. 
Note that $\erfc(z)=\frac{2}{\sqrt{\pi}}\int_z^\infty \exp(-t^2)\mathrm{d}t$ is the complementary error function. 
The proof of \eqref{e4} is included in Appendix \ref{app:A}.
Further note that,  the fraction of IMs absorbed within time $t$ by the FAR$_i$ (denoted by $ \pone{i}{t,a,r_i} $) in the absence of any other FAR is  \cite{Yilmaz2014}	
		\begin{align}
		\pone{i}{t,a,r_i}=\frac{a}{r_i}
		\erfc\left(\frac{r_i-a}{\sqrt{4Dt}}\right).\label{e3}
		\end{align}
			 \begin{figure*}[t!]
			\centering
			\includegraphics[width=0.32\linewidth]{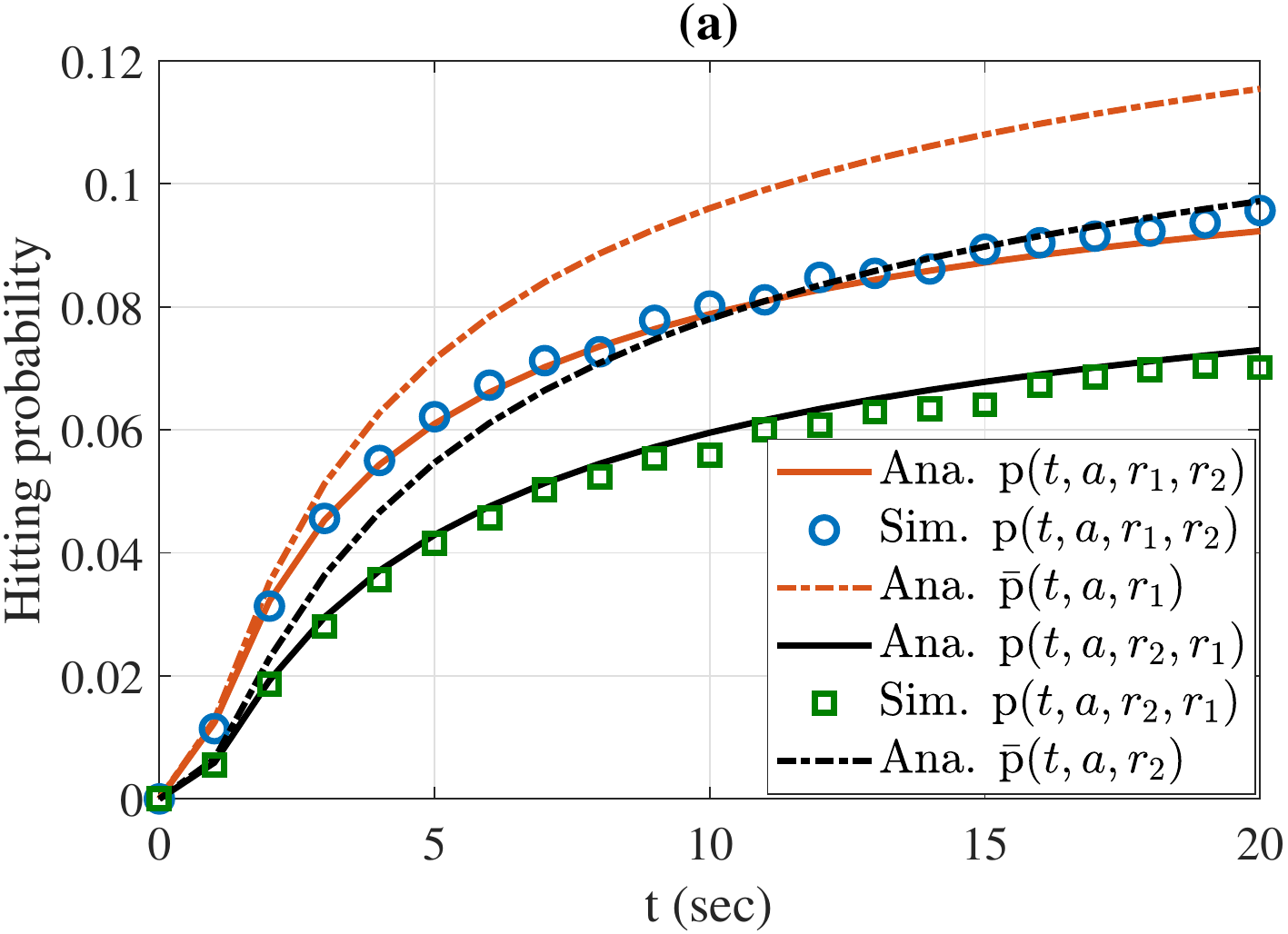}	
			\includegraphics[width=0.32\linewidth]{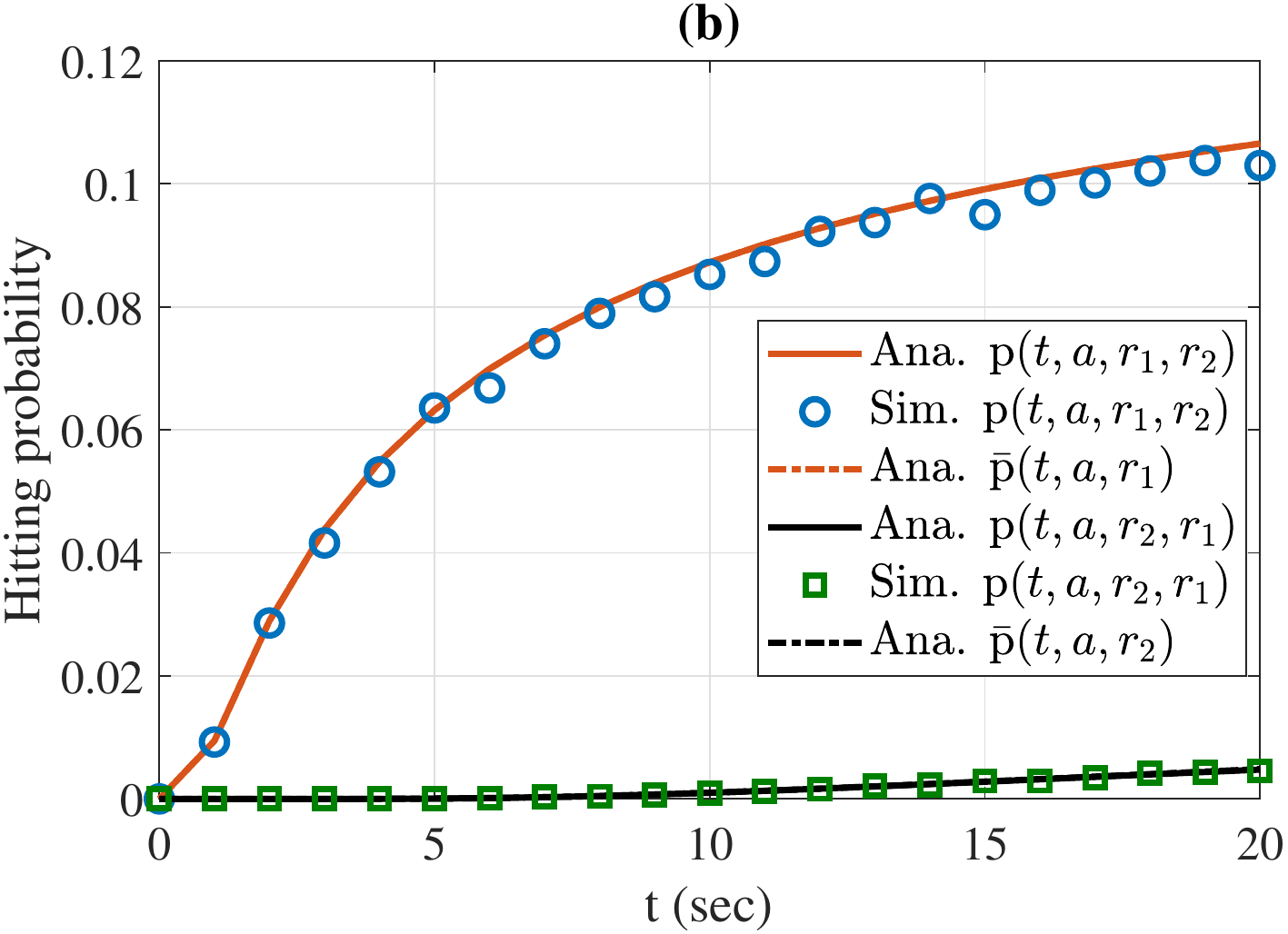}	
			\includegraphics[width=0.32\linewidth]{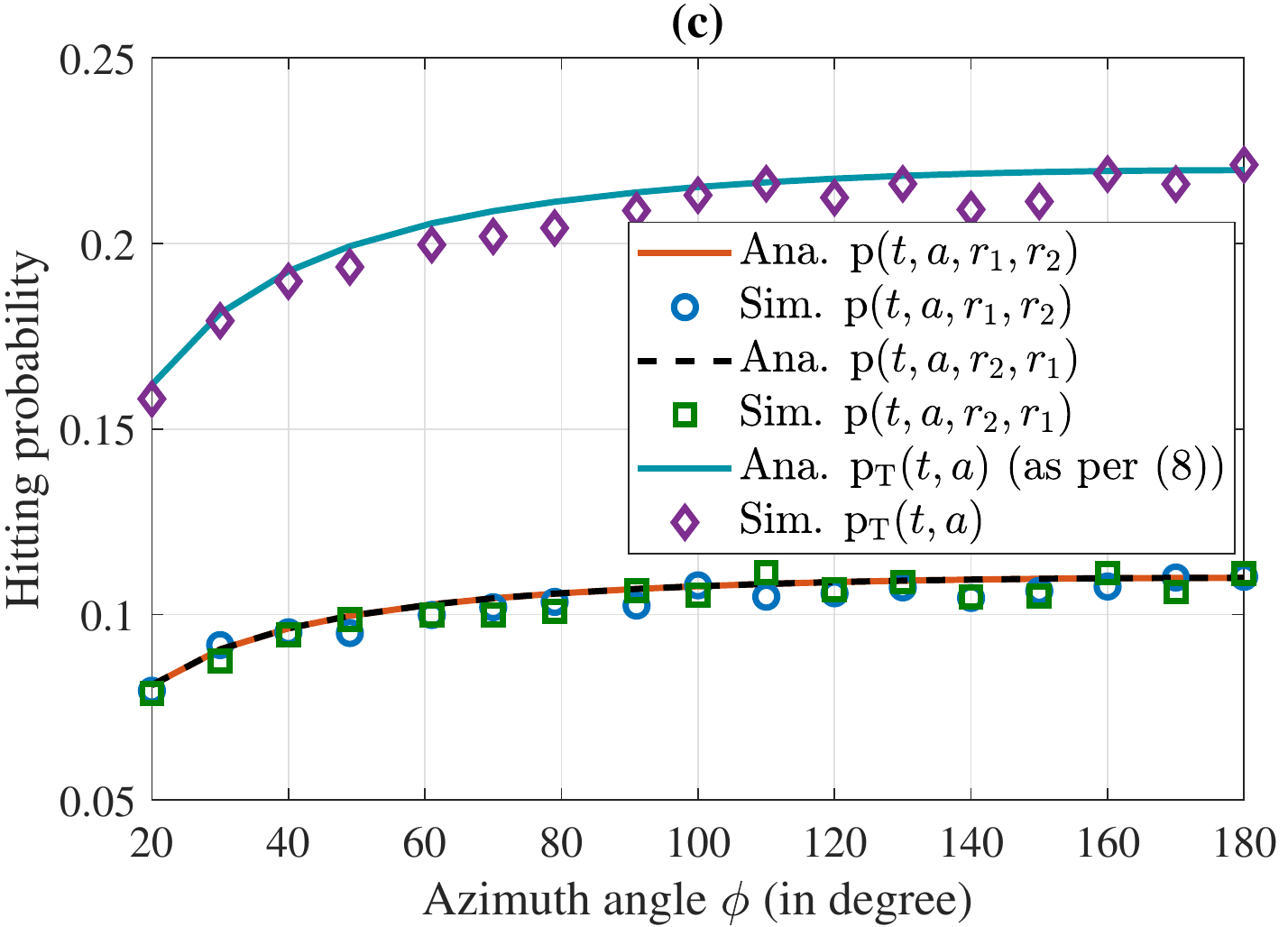} 
			\vspace{-0.2in}		
			\caption {The variation of hitting probability $\ptwo{i}{t,a,r_i,r_j}$. (a) $\ptwo{i}{t,a,r_i,r_j}$ versus time $ t $ when $\x_1=[30,0,0]$ and $\x_2=[30,15,0]$. (b) $\ptwo{i}{t,a,r_i,r_j}$ versus $ t $ when $\x_1=[-30,-10,0]$ and $\x_2=[100,40,0]$. (c) $\ptwo{i}{t,a,r_i,r_j}$ versus the angular distance $\phi$ between the two FARs at time  $ t=15$s. The minimum azimuth angle is taken as $20^\mathrm{o}$ to avoid the overlap of FARs.
				The solid/dashed lines represent the analytical expression  (terms  $< 10^{-16} $ are neglected), and markers represent the values obtained via particle-based simulations. Here, analytical values of $\ptwo{i}{t,a,r_i,r_j}$ and $\pone{i}{t,a,r_i}$ are as per \eqref{e4} and \eqref{e3} respectively.}
			\vspace{-0.5cm}
			\label{fig3}
		\end{figure*}\par
 From \eqref{e4} and \eqref{e3}, we can develop several important insights that are presented below. 	
		\begin{coro}\label{corr1}
			The fraction of IMs eventually hitting FAR$ _i $ in the absence and the presence of FAR$ _j $ are
			\begin{align}
			\pone{i}{\infty,a,r_i}&=\frac{a}{r_i} \text{, and }
			\ptwo{i}{\infty,a,r_i,r_j}= \frac{aR_{ij}}{R_{ij}R_{ji}-a^2} \left[\frac{R_{ji}}{r_i}{-}\frac{a}{r_j}\right].\label{e5}
			\end{align}
			Therefore, the presence of FAR$_j$ reduces the  eventual hitting probability at FAR$_i$ by 
			the amount 
			\begin{align}
				\pred{i}{\infty,a}&=	\pone{i}{\infty,a,r_i}-\ptwo{i}{\infty,a,r_i,r_j}=
				\frac{a^2}{R_{ij}R_{ji}-a^2} \left[\frac{R_{ij}}{r_j}{-}\frac{a}{r_i}\right],\label{eq9}
			\end{align}
			which denotes the fraction of IMs that would have hit FAR$_i$ eventually, 
			but instead hit FAR$_j$ first and got absorbed. 
		\end{coro}
	\begin{coro}\label{corr2}
		When the two FARs are far apart, (i.e. $R\rightarrow \infty$. Note that, $R\rightarrow \infty \implies R_{ij},R_{ji}\rightarrow \infty$),
		\begin{align}
		\ptwo{i}{t,a,r_i,r_j}\rightarrow \frac{a}{r_i}\erfc\left(\frac{r_i-a}{\sqrt{4Dt}}\right)=\pone{i}{t,a,r_i}.\label{e7}
		\end{align}	
		In other words, the mutual influence of FARs vanishes as they move away from each other.
	\end{coro}
	\begin{coro}\label{corr3}
		The probability that an IM reaches any of the FARs is
		\begin{align}
		\ptotal{i}{t,a}=\ptwo{1}{t,a,r_1,r_2}+\ptwo{2}{t,a,r_2,r_1}.\label{eq:ptot}
		\end{align}
	\end{coro}
\subsection{Validation}
We first validate the expression \eqref{e4}  of $\ptwo{i}{t,a,r_i,r_j}$   through  particle-based simulations which are carried out for $10^4$ iterations,  with a step size of $10^{-4}$\,s.   $D$ is  $100\ \mu$m$/$s$^2$, and both FARs have radius $a=5\,\mu$m, which are the same for all numerical evaluations in this paper  unless stated otherwise. Fig. \ref{fig3} (a) and (b) show the hitting probability of IM on each of the FARs in presence of the other for two different cases.  Fig. \ref{fig3} (a) shows the variation of hitting probability with $ t $ when $\x_1=[30,0,0]$ and $\x_2=[30,15,0]$ and Fig. \ref{fig3} (b) shows the variation of hitting probability with $ t $ when $\x_1=[-30,-10,0]$ and $\x_2=[100,40,0]$. In the Fig. \ref{fig3} (a),  the FARs are relatively closer to each other with $R=15\microm$. In Fig. \ref{fig3} (b),  the two FARs are relatively far away with $R=139.2\microm$.
We can observe that the analytical expression \eqref{e4}  closely matches with simulation results for both cases.  We can also observe that in Fig. \ref{fig3} (a), FARs have a significant influence on each other, which 
grows  with time $t$  as seen by the widening gap between solid and dashed lines. In Fig. \ref{fig3} (b), 
the distance between FARs is large enough, resulting in a negligible  mutual influence.
Also, the hitting probability of the FAR closer to the transmitter of the two FARs is higher than that of the other one.

From extensive numerical simulations, we found that the absolute error ($ |\text{analytical approximate}$  value $-\ \text{exact value}| $) of the hitting probability expression for each FAR is negligible when $ r_1{>}3a,\ r_2{>}3a $ and $ R{>}3a $ implying the  goodness of approximation under these conditions. Note that, the conditions $r_1,\ r_2{>}a$ and $ R{>}2a $ is a prerequisite to avoid the overlap between FARs and FAR and transmitter.
\subsection{Impact of  Distance $R$ on Hitting Probability}
We now study the impact of mutual distance on the hitting probability of an IM on FARs equidistant from the transmitter. Without loss of generality, we consider one FAR at \xaxis-axis with $r_1{=} 20\microm$ and other FAR in \xaxis-\yaxis-plane with the same radial distance $\norm{\x_2}{=}r_1$ and azimuth angle $\phi$. Note that, the distance depends on $\phi$ as 
$R{=}2\norm{\x_2}\sin{(\phi/2)}$.
Fig. \ref{fig3} (c) shows the variation of hitting probability with varying azimuth angular distance $\phi$ between the two FARs.
Here also, we can observe that for the chosen parameters, the analytical and simulation results match well, including the scenario when the FARs are close to each other. 
Fig. \ref{fig3} (c) also shows the total probability $ \ptotal{}{t,a}$. It can be verified that  $ \ptotal{}{t,a}=2\ptwo{1}{t,a,r_1,r_2}$. This is because the fraction of IMs absorbed by each FARs are the same owing to their equal distance from the transmitter. 
\subsection{Comparison of two FARs vs single FAR}
We now study the gain $g(t,a)$ that can be achieved by replacing one FAR  by two FARs  at two different locations.
In particular, in the first scenario, there is only one FAR  of radius $a$ at $ \x_1$ with hitting probability of IM as $\pone{1}{t,a,r_1}$. Now, in the second scenario, there are two  FARs, each of radius $b$ at two different locations $ \x_1$ and $\x_2$ such that $r_1=r_2$. 
For a fair comparison, we keep the total surface area of the  FARs equal in both scenarios \ie $b{=}a/\sqrt{2}$. The hitting probability of an IM on any of the FARs is $\ptotal{}{t,a/\sqrt{2}}{=}2\ptwo{1}{t,a/\sqrt{2},r_1,r_2}$. We can see that for any time $t$,
		\begin{align}
		\!\!		
		g(t,a)=\frac{\ptotal{}{t,a/\sqrt{2}}}{\pone{1}{t,a,r_1}}&<
		\frac
		{\sqrt{2}\erfc\left(\frac{r_1-a/\sqrt{2}}
		{\sqrt{4Dt}}\right)}
		{\erfc\left(\frac{r_1-a}{\sqrt{4Dt}}\right)}
		<{\sqrt{2}}\label{eqn12}
		\end{align}
which upper bounds the gain.
Further, using the following lower  and upper bounds \cite{erfcul} of $\erfc$: 
\begin{center}
$\frac{e^{-x^2}}{\sqrt{\pi}x}\left(1-\frac{1}{2x^2}\right)$ $<\erfc(x)$ $<\frac{e^{-x^2}}{\sqrt{\pi}x}$
\end{center}
 in  the denominator and numerator terms of $g(t,a)$, we can show that, for any $ t $, 
\begin{align}
	\!\!\!\!	
	g(t,a)
	<
	&\frac{r_1-a}{r_1-\frac{a}{\sqrt{2}}}
	\frac{\exp\left(-{\frac{
	(2-\sqrt{2})
	ar_1-a^2/2}{4Dt}}\right)}
	{\left(\frac{1}{\sqrt{2}}{-}\frac{\sqrt{2}Dt}{\left(r_1-a\right)^2}\right)}
	,\label{e14}
\end{align}	
which is less than 1 for small $t$. This implies that for small $t$, the scenario with a single FAR gives better hitting probability.

\begin{figure}
\centering
\includegraphics[width=0.6\linewidth]{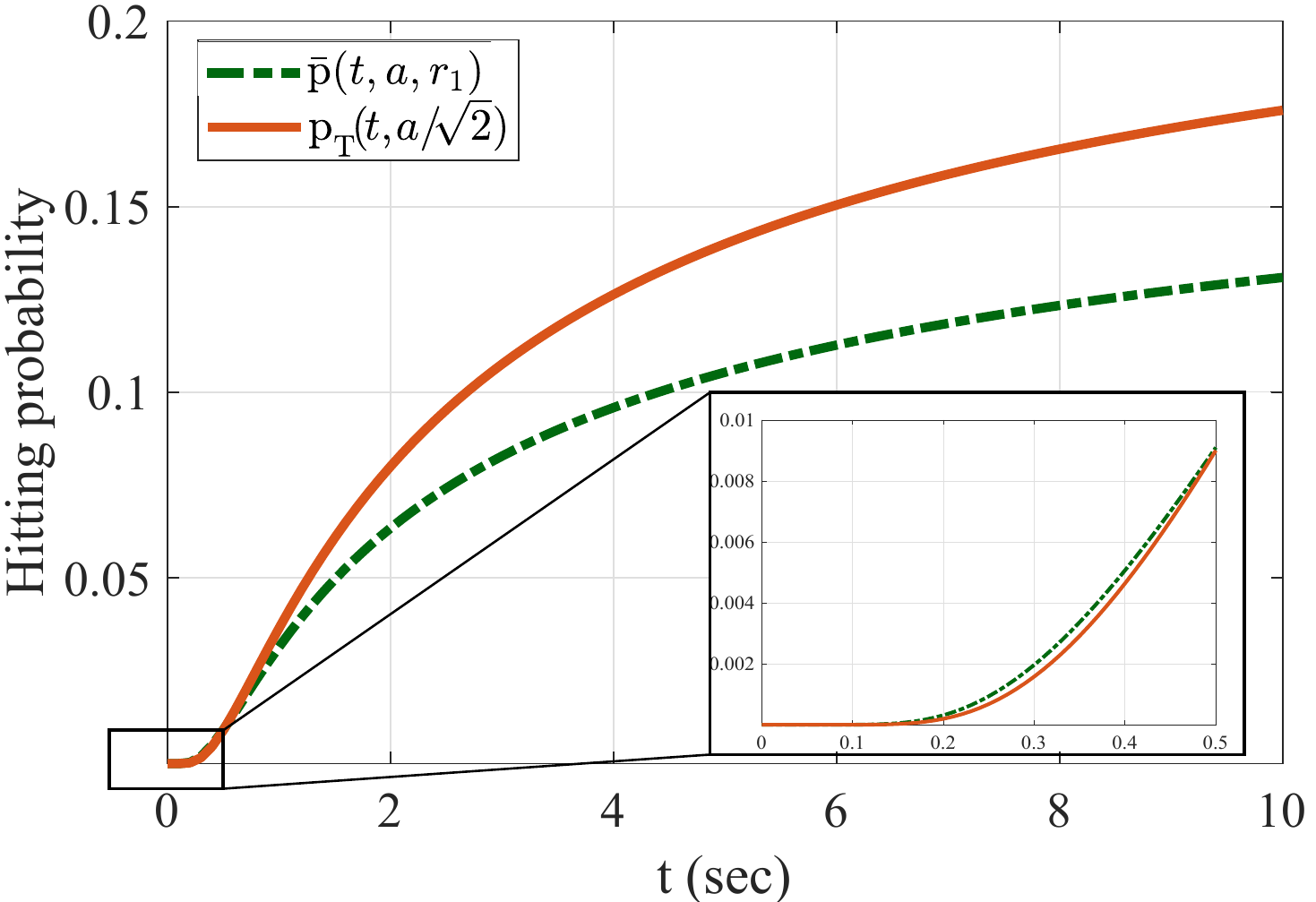}  
\vspace{-0.2cm}
\caption{Comparison of the hitting probability of IMs for the case with a single FAR of 
  radius $ a$ located
  at $ \x_1=[25 \ 0\ 0] $ vs the case with two receivers of radius $ a/\sqrt{2} $ at
  $ \x_1=[25 \ 0\ 0] $ and $ \x_2=[-25\ 0\ 0] $. Here, $a=5\mu$m.
  }
\vspace{-0.6cm}
\label{split}
\end{figure}

%
However, when $ t \rightarrow \infty$ and 
$R_{ij}>a(1+1/\sqrt{2})$,
 \eqref{e5} gives
		\begin{align}
		g(\infty,a)=\frac{\ptotal{i}{\infty,a/\sqrt{2}}}{\pone{1}{\infty,a,r_1}}=
		\frac{2R_{ij}}{\sqrt{2}R_{ij}+a}
		>1,\label{e13}
		\end{align}	
which implies that the  hitting probability of an IM on any of the FARs of radius $a/\sqrt{2}$ is higher than 	the scenario with a single FAR of radius $a$. Fig. \ref{split}  compares the two scenarios. It can be seen that initially, the single FAR gives better hitting probability.  This is due to the close proximity of the surface of a single FAR of radius $ a $ than two FARs of radius $ a/\sqrt{2} $ to the transmitter. However, as time $ t $ increases, the total hitting probability of IM on any one of the two FARs becomes larger, which is consistent with the above analysis.  This increase in hitting probability is due to the absorption of IM in more directions by the two FAR case compared to that of a single FAR case.

\section{Detection Performance at FARs}\label{sec:auc}
Let $\pdetect{i}{l}$ and $\pfalse{i}{l}$ denote  the  detection and false alarm probabilities at FAR$ _i $ in the $ l $th time-slot, respectively. 
Applying the binary hypothesis testing \cite{bht}  on $Y_i[l]$ for the detection of bit $b[l]$, $\pdetect{i}{l}$ and $\pfalse{i}{l}$ can be derived from  \eqref{RecMol} as
\begin{align}
\!\!\!\!	
\pdetect{i}{l}=
&\pro\left[Y_i[l]{>}\eta_i\mid b[l]{=}1\right]=
Q\left(\frac{\eta_i-\mu_1[i;l]}{\sigma_1[i;l]}\right), \label{Pd_N}\\
\!\!\!\!	
\pfalse{i}{l}=
&\pro\left[Y_i[l]{>}
\eta_i\mid b[l]{=}0\right]=
Q\left(\frac{\eta_i-\mu_0[i;l]}{\sigma_0[i;l]}\right). 
\label{Pf_N}
\end{align}
Here, $ Q(x)=0.5\erfc\left(x/\sqrt{2}\right) $ is the standard $Q$-function. 

The receiver operating characteristic (ROC) curve illustrates the variation of the detection probability with respect to the false alarm probability for a receiver by varying detection threshold as an intermediate variable. 
The area under the ROC curve (AUC) is a quantitative measure of a receiver's capability of correct decision 
\cite{Atapattu2010}. The AUC value can vary from $0$ to $1$, where AUC = 0  indicates that bits are always erroneously decoded, and AUC = 1 indicates the perfect decoding without any error. Moreover, AUC = 0.5 indicates that the receiver is unable to distinguish between 0 and 1.
The AUC for the FAR$_i$ in $l$th time-slot is \cite{Atapattu2010}
	\begin{align}
		\auc{i}{l} &= 
		\int_0^1 \pdetect{i}{l} \dd \pfalse{i}{l}.
		\label{AUC1}\\
		&{=}\frac{1}{\sqrt{2\pi}\sigma_0[i;l]}\int_0^\infty Q\left(\frac{\eta_i-\mu_1[i;l]}{\sigma_1[i;l]}\right)\times\exp\left(-\frac{(\eta_i-\mu_0[i;l])^2}{2\sigma_0^2[i;l]} \right) \mathrm{d}\eta_i.\label{eq26}
	\end{align}
	
Applying approximation on $Q$-function, a closed-form approximation 
for AUC at FAR$_i$ in $l$th time-slot is derived as 
%
%
%
\begin{align}
\!\!\!\!	
\auc{i}{l}
&\approx\frac{1}{2\sqrt{2}\sigma_0[i;l]}\sum_{\kappa=1}^3\frac{(-1)^{\kappa+1}}{\sqrt{a_\kappa}}\exp\left(\frac{b_\kappa^2-a_\kappa c_\kappa}{a_\kappa}\right)\nonumber\\
	&\times
	\!\!
	\left[
		\erfc\!\left(\!\!e_\kappa{+}\frac{b_\kappa}{\sqrt{a_\kappa}}\right)
		-\erfc\!\left(
			\frac{\sqrt{a_\kappa}}{d_\kappa\mu_1[i;l]} 
			{+}
			\frac{b_\kappa}{d_\kappa\sqrt{a_\kappa}}
		\right)
	\right]\label{approx}\!,\!\!
	\end{align}
\noindent \hspace{-.1in} where $ \alpha{=} 0.3842$, $ \beta{=} 0.7640$, $ \gamma{=}0.6964 $, 
$a_1=0.5 \sigma_0^{-2}$, 
$b_1=-\mu_0a_1$, 
$c_1=\mu_0^2a_1$, 
$d_1=1$,
$e_1=0$, 
$a_2={\alpha}\sigma_1^{-2}+a_1$, 
$b_2=-(\alpha\mu_1+\beta\sigma_1/2)\sigma_1^{-2}+b_1$, 
$c_2={(\alpha\mu_1+\beta\sigma_1)\mu_1}{\sigma_1^{-2}}+c_1+\gamma$,  
$d_2=1$, 
$e_2=0$, 
$a_3=a_2$, 
$b_3=-({\alpha\mu_1-\beta\sigma_1/2}){\sigma_1^{-2}}+b_1$, 
$c_3={(\alpha\mu_1-\beta\sigma_1)\mu_1}{\sigma_1^{-2}}+c_1+\gamma$, 
$d_3=0$, 
$e_3=\sqrt{a_3}\mu_1$
for respective $i$ and $l$.

\begin{figure}[t!]
		\centering
		\includegraphics[width=0.7\linewidth]{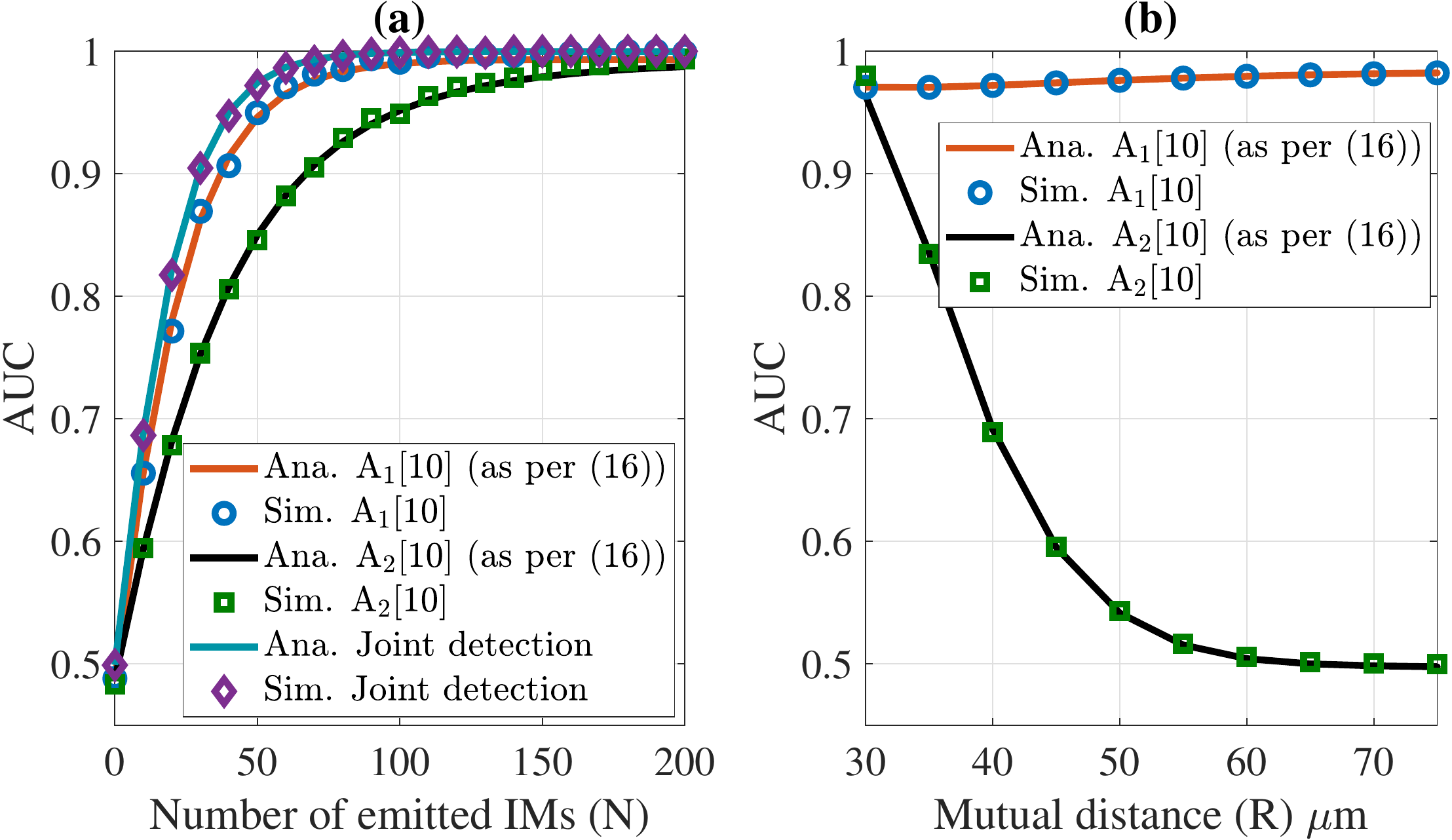}  
		\caption{(a) AUC of the individual FARs vs $N$ (number of emitted IMs). Here, $\x_1=[20,5,0]$ and $\x_2=[-25,-10,0]$. 
		(b) Impact of mutual distance $R$ between the two FARs on their AUC. 
		For both figures, $\mu_n=\sigma_n^2=5$.  The solid lines represent the analytical values obtained using \eqref{approx}, whereas the markers represent the values obtained using  Monte-Carlo simulations.  
		}
		\label{AUCf}
		\vspace{-0.5cm}
	\end{figure}

Fig. \ref{AUCf}(a) shows the AUC variation with $N$ for both FARs in $l{=}10$th slot for $\ts=5$s. 
It can be  observed that the AUC values 
at both FARs significantly improve as $N$ increases.   This improvement in AUC 
 is due the fact that the gap between $ Y_i[l] $ for $ b[l]=1 $ and  $ b[l]=0 $ increases with $N$ with respect to the noise $N_i[l]$, and the variance of $S_i[l]$ does not increases relatively as much as its mean with $N$. Further we can observe that, the receiver closer to the transmitter, which is  FAR$_1$ here, has larger  AUC  than FAR$_2$ which indicates better decision capability of FAR$_1$.  We also study a scenario with two FARs working together to make a joint detection. In this case,  $ Y[l] =Y_1[l] +  Y_2[l] $ is compared with threshold $ \eta $ to make a decision for bit $b[l]$. The AUC of this joint detection is given by \eqref{eq26} with the mean  and variance values as $ \mu_0[l]=\mu_0[1;l]+\mu_0[2;l] $, $ \mu_1[l]=\mu_1[1;l]+\mu_1[2;l] $, $ 	\sigma_0^2[l] =	\sigma_0^2[1;l] +	\sigma_0^2[2;l] $ and $ 	\sigma_1^2[l] =	\sigma_1^2[1;l] +	\sigma_1^2[2;l] $ respectively.   Fig. \ref{AUCf}(a) also shows the AUC of this system. It can be seen that its AUC  is higher than individual AUC of both FARs which is intuitive. An ideal implementation of such system would require a central node which is transparent to IMs for combining the observation from FARs. Hence, these AUC values serve as the upper bound to any practical implementation.

	Fig. \ref{AUCf}(b) shows variation in the AUC  with  distance between the two FARs. 
	Here, $N=1000$ and $\ts=1$s. 
	FAR$_1$ is fixed at $\x_1=[{-}10,0,0]$. 
	The location of the FAR$_2$ is $\x_2=[-10+R,0,0]$ which is  moved in positive \xaxis-direction by increasing $R$.
	It can be seen that the increment in $R$ results in higher distance between transmitter and FAR$_2$, which in turn deteriorates the detection capability at FAR$_2$. It is interesting to note that even though influence of FAR$_1$ on FAR$_2$ reduces with increase in $R$, the gain in the number of received IMs at FAR$_2$ is superseded by the loss of IMs due to increase in distance of FAR$_2$ from the transmitter. The performance at FAR$_1$ improves slightly since the number of IMs reaching FAR$_1$ increases due to the diminishing influence of FAR$ _2 $ on FAR$ _1 $ with increase in  $ R $. One can also note that both FARs have identical AUC values when they are located at an equal distance from the transmitter.

\section{Conclusions}
For a 3D MCvD system with multiple FARs, there is no analytical channel model in the current literature. 
In this work, we have tried to bridge this gap by presenting an approximate analytical expression for hitting probability of an IM considering two FARs in $ \rthree $ space. We have developed several important insights that are lacking in the current literature.  Moreover, this work explicitly demonstrated the impact of receiver locations on their mutual dependency.  We have found that the use of two distantly located receivers can increase the total hitting probability by covering two different directions of molecular movement from the transmitter compared to the use of single FAR in one direction. Using the hitting probability expression, this work analyzed the detection performance at both FARs in terms of AUC and quantified the impact of their location on their detection capability. 
Future work can now focus on (a) characterizing the 3D channel for more than two FARs by applying similar techniques discussed in this work, and (b) applying presented results in the analysis of large scale networks.
	
	\appendices
\section{Derivation of \eqref{e4}}\label{app:A}
The probability that an IM emitted by the point source at origin hits the surface of FAR$ _j $ in the interval $ [\tau,\ \tau+\dd\tau] $ is 
$ \left[\frac{\dd \ptwo{i}{t,a,r_j,r_i}}{\dd\tau}\right]\dd \tau $. 
The probability that this IM hits the FAR$ _i $ in the remaining $ t-\tau $ time is $  \pone{i}{t-\tau,a,d_{ji}} $, where $d_{ji}$ is the distance between the IM's hitting point at the surface of FAR$_j$ and the center of FAR$_i$. Note that $d_{ji}$ is a random variable. To simplify the analysis, we approximate the term $ \pone{i}{t-\tau,a,d_{ji}} $ by $  \pone{i}{t-\tau,a,R_{ji}} $, where $ R_{ji} $ is the distance between $E_j$ (the nearest point on the surface of FAR$ _j $ from the transmitter) and the center of FAR$ _i $ (see Fig. \ref{fig:sm}). 
The probability of an IM that is supposed to hit the FAR$ _i $ within time $ t $, but gets absorbed at FAR$ _j $ before hitting FAR$ _i $ is \cite[Eq. 13]{Berezhkovskii1990}\cite[Eq. A2]{Berezhkovskii1989}
\begin{align}
\pone{i}{t,a,r_i}{-}& \ptwo{i}{t,a,r_i,r_j}{=}{\int_{0}^{t}}\frac{\partial \ptwo{j}{t,a,r_j,r_i}}{\partial\tau}\pone{i}{t{-}\tau,a,R_{ji}}\dd\tau.\label{ae1}
\end{align}
Similarly, the probability of an IM that is supposed to hit the FAR$ _j $ within time $ t $, but is hitting the FAR$ _i $ before it,  is
\begin{align}
\pone{j}{t,a,r_j}{-}& \ptwo{j}{t,a,r_j,r_i}
{=}{\int_{0}^{t}}\frac{\partial \ptwo{i}{t,a,r_i,r_j}}{\partial\tau}\pone{j}{t{-}\tau,a,R_{ij}}\dd\tau.\label{ae2}
\end{align}
Now, taking the Laplace transform (LT) of \eqref{ae1} and \eqref{ae2} gives
\begin{align}
\lpone{i}{s,a,r_i}{-} \lptwo{i}{s,a,r_i,r_j}{=}s\lptwo{j}{s,a,r_j,r_i}\lpone{i}{s,a,R_{ji}},\label{ae3}\\
\lpone{i}{s,a,r_j}{-} \lptwo{j}{s,a,r_j,r_i}{=}s\lptwo{i}{s,a,r_i,r_j}\lpone{j}{s,a,R_{ij}},\label{ae4}
\end{align}
where $ \lpone{i}{s,a,r_i},\ \lptwo{i}{s,a,r_i,r_j}$ and $\lpone{i}{s,a,R_{ij}}$ are the LTs of $ \pone{i}{t,a,r_i},\ \ptwo{i}{t,a,r_i,r_j}$ and $\pone{i}{t,a,R_{ij}}$, respectively.\\
  Solving \eqref{ae3} and \eqref{ae4} gives 
 \begin{align}
 \lptwo{i}{s,a,r_i,r_j}=&\frac{\lpone{i}{s,a,r_i}-s\lpone{j}{s,a,r_j}\lpone{i}{s,a,R_{ji}}}{1-s^2\lpone{i}{s,a_,R_{ij}}\lpone{i}{s,a_,R_{ji}}}\label{ae5},
 \end{align}
where  $ \lpone{i}{s,a,x}$ can be solved as
 \begin{align*}
 \lpone{}{s,a,x}=\mathcal{L}\left[\frac{a}{x}\erfc\left(\frac{x-a}{\sqrt{4Dt}}\right)\right]=\frac{a}{x}\frac{\exp\left(-\left(x-a\right)\sqrt{\frac{s}{D}}\right)}{s}.\label{eq:app1:laplaceP}
 \end{align*}
Finally, substituting the above expression in \eqref{ae5} and taking the inverse LT  gives \eqref{e4}.\newpage

\begin{center}
\LARGE{Supplementary file to `3-D Diffusive Molecular Communication with Two Fully-Absorbing Receivers: Hitting Probability and Performance Analysis'}
\end{center}
	\section{Derivation of mean $ \mu_{b[l]}[i;l]$ and variance $\sigma^2_{b[l]}[i;l] $ of $Y_i[l]$}

\noindent From  \eqref{RecMol}, we know
\begin{align}
	Y_i[l] &= S_i[l]+\sum_{k=1}^{l-1}I_i[k]+N_i[l]
\end{align}
with 
\begin{align*}
	S_i[l]&\sim\mathcal{N}(Nb[l]h_i[0],Nb[l]h_i[0](1-h_i[0]))\\
	I_i[k]&\sim\mathcal{N}\left(Nb[k]h_i[l-k],Nb[k]h_i[l-k](1-h_i[l-k])\right)\\
	N_i[l]&\sim\mathcal{N}(\mu_{\mathrm{n}},\sigma^2_{\mathrm{n}})
	.
\end{align*}
Also for time-slot $k<l$, the transmit bit $b[k]$ is an independent Bernoulli random variable taking value 1 with probability $q_1$, and  $0$ with $q_0=1-q_1$. Hence $\expect{}{b[k]}=q_1$.

Now, given the bit transmitted at the current slot $ l $ \ie  $ b[l] $, 
the mean of the random variable $ Y_i[l] $ is given 
\begin{align}
	\mu_{b[l]}[i;l]&=\expect{}{Y_i[l]}\nonumber\\&=Nh_i[0]b[l]+\mathbb{E}\left[\sum_{k=1}^{l-1}Nb[k]h_i[l{-}k]\right]+\mu_\mathrm{n}\nonumber\\
	&=Nh_i[0]b[l]+Nq_1\sum_{k=1}^{l-1}h_i[l{-}k] + \mu_\mathrm{n}.
\end{align}
For the derivation of variance of $ Y_i[l] $, first, we derive the variance of $ I_i[k] $. The mean of $ I_i[k] $ given $ b[k]$ is
\begin{align}
	\expect{}{ I_i[k]\mid b_i[k] }&= Nb[k]h_i[l-k].\label{v1}
\end{align}
The variance of $ I_i[k] $ given $ b[k] $ is
\begin{align}
	\var{ I_i[k]\mid b[k]}&=Nb[k]h_i[l{-}k](1{-}h_i[l{-}k]).\label{v2}
\end{align}
From \eqref{v1} and \eqref{v2}, the variance of  $ I_i[k] $ can be derived as 
\begin{align}
	\var{ I_i[k]})=&\expect{}{\mathrm{Var}( I_i[k]\mid b_i[k])}+\mathrm{Var}(\expect{}{ I_i[k]\mid b_i[k] })\nonumber\\
	=&Nq_1h_i[l{-}k](1{-}h_i[l{-}k])+N^2h_i[l-k]^2\var{b[k]}\nonumber\\
	=&Nq_1h_i[l{-}k](1{-}h_i[l{-}k])+N^2h_i[l-k]^2q_1q_0
\end{align} 
Therefore, variance of $ Y_i[l] $ given the current transmitted bit $b[l]$ is
\begin{align}
	\sigma_{b[l]}^2[i;l] =&\mathrm{Var}(S_i[l])+\mathrm{Var}(\ISI_i[l])+\mathrm{Var}(N_i[l])  \nonumber\\ =&Nb[l]h_i[0](1{-}h_i[0]){+}Nq_1\sum_{k=1}^{l-1}\left[h_i[l{-}k](1{-}h_i[l{-}k])\right.\nonumber\\
	&\left.+Nq_0h_i[l{-}k]^2\right]+\sigma_n^2.
\end{align}

\section{Derivation of \eqref{e14}}

\noindent Using \eqref{e4} and \eqref{e3}, for any $ t $,
\begin{align}	g(t,a)=\frac{\ptotal{}{t,a/\sqrt{2}}}{\pone{1}{t,a,r_1}}&<
	\frac
	{\sqrt{2}\erfc\left(\frac{\norm{\x_1}-a/\sqrt{2}}
		{\sqrt{4Dt}}\right)}
	{\erfc\left(\frac{\norm{\x_1}-a}{\sqrt{4Dt}}\right)}.\label{a7}
\end{align}
Now, using the following upper  and lower bounds \cite{erfcul} of $\erfc$: 
\begin{center}
	$\frac{e^{-x^2}}{\sqrt{\pi}x}\left(1-\frac{1}{2x^2}\right)$ $<\erfc(x)$ $<\frac{e^{-x^2}}{\sqrt{\pi}x}$,
\end{center}
at the numerator and the denominator respectively of \eqref{a7} gives
\begin{align}	\frac{\ptotal{i}{t,a/\sqrt{2}}}{\pone{1}{t,a,r_1}}
	&<\frac{\frac{\sqrt{2}{\exp\left(-\left(\frac{\norm{\x_1}-a/\sqrt{2}}{\sqrt{4Dt}}\right)^2\right)}}{\sqrt{\pi}\left(\frac{\norm{\x_1}-a/\sqrt{2}}{\sqrt{4Dt}}\right)}}{\frac{{\exp\left(-\left(\frac{\norm{\x_1}-a}{\sqrt{4Dt}}\right)^2\right)}}{\sqrt{\pi}\left(\frac{\norm{\x_1}-a}{\sqrt{4Dt}}\right)}\times\left(1-\frac{1}{2\left(\frac{\norm{\x_1}-a}{\sqrt{4Dt}}\right)^2}\right) }\label{eq:app11}.
\end{align}
Simplifying \eqref{eq:app11} 
gives
\begin{align}
	\!\!\!\!	
	\frac{\ptotal{i}{t,a/\sqrt{2}}}{\pone{1}{t,a,r_1}}
	<
	&\frac{\norm{\x_1}-a}{\norm{\x_1}-\frac{a}{\sqrt{2}}}
	\frac{\exp\left(-{\frac{
				{(2-\sqrt{2})}
				a\norm{\x_1}-a^2/2}{4Dt}}\right)}
	{\left(\frac{1}{\sqrt{2}}{-}\frac{\sqrt{2}Dt}{\left(\norm{\x_1}-a\right)^2}\right)}
	,
\end{align}
which is \eqref{e14} in the submitted manuscript.

\section{Derivation of \eqref{e13}}

\noindent The inequality shown in \eqref{e13} can be derived from \eqref{e5} as
\begin{align}
	g(t,a)=\frac{\ptotal{i}{\infty,a/\sqrt{2}}}{\pone{1}{\infty,a,r_1}}&=\frac{\frac{2\times a/\sqrt{2}}{\norm{\x_1}} \times \frac{R_{ij}}{R_{ij}+a/\sqrt{2}}}{\frac{a}{\norm{\x_1}}}\nonumber\\
	&=\sqrt{2} \times \frac{R_{ij}}{R_{ij}+a/\sqrt{2}}\nonumber\\
	&=\frac{2R_{ij}}{\sqrt{2}R_{ij}+a}\label{a5}
\end{align}	
Note that, $ R>a $ for transmitter to not to overlap with the FAR and $ R\gg a $ ($ R\gg a \implies R_{ij}\gg a$) for \eqref{e4} to be valid with minimum error. When $ R_{ij}>a\left(1+1/\sqrt{2}\right) $, \eqref{a5} is
\begin{align}
	g(\infty,a)=\frac{\ptotal{i}{\infty,a/\sqrt{2}}}{\pone{1}{\infty,a,r_1}}=
	\frac{2R_{ij}}{\sqrt{2}R_{ij}+a}>1,
\end{align}
which is \eqref{e13} in the submitted manuscript.

\section{Derivation of \eqref{approx}}\label{app:B}
\noindent The probability of false alarm $\pfalse{i}{l}$ varies monotonically from 0 to 1 when  $\eta_i$ changes from $\infty$ to 0. Thus, \eqref{AUC1} can be equivalently written as \cite{Atapattu2010}
\begin{align}
	\text{A}_i[l] = -\int_0^\infty \pdetect{i}{l}\frac{\mathrm{d}\pfalse{i}{l}}{\mathrm{d}\eta_i} \mathrm{d}\eta_i,     \label{eqAUC}
\end{align}
where
\begin{align}
	\!\!	\frac{\mathrm{d}\pfalse{i}{l}}{\mathrm{d}\eta_i} = -\frac{1}{\sqrt{2\pi}\sigma_0[i;l]}\exp\left(-\frac{(\eta_i-\mu_0[i;l])^2}{2\sigma_0^2[i;l]} \right).
\end{align}
Further, substituting the above expression along with (13) in (17), the A$_i[l]$ can be written~as
\begin{align}
	\!\!\!	\text{A}_i[l] 
	{=}\frac{1}{\sqrt{2\pi}\sigma_0[i;l]}&\int_0^\infty Q\left(\frac{\eta_i-\mu_1[i;l]}{\sigma_1[i;l]}\right)\nonumber\\&\times\exp\left(-\frac{(\eta_i-\mu_0[i;l])^2}{2\sigma_0^2[i;l]} \right) \mathrm{d}\eta_i.\label{eq26}
\end{align}
Finally, splitting the above integral into two separate integrals with limits from $0$ to $\mu_1[i;l]$ and from $\mu_1[i;l]$ to $\infty$,     	and subsequently using the following tight and more tractable approximation\footnote{The fitting coefficients for positive and negative argument are optimized to minimize the sum of square errors. } for $Q(x)$ \cite{Lopez-Benitez2011}
\begin{align}
	Q(x)\approx\begin{cases}
		\exp(-\alpha x^2-\beta x-\gamma) ~&\text{if}~ x\geq 0\\
		1-\exp(-\alpha x^2+\beta x-\gamma)~ &\text{if}~ x< 0,
	\end{cases}\label{exp}
\end{align}
(where $ \alpha= 0.3842$, $ \beta= 0.7640$ and $ \gamma=0.6964 $) and then, using the following integral identity from \cite[Eq. 2.33.1]{jeffrey2007table}, \ie
\begin{align*}
	\int e^{-\left(ax^2+2bx+c\right)}\mathrm{d}x{=}\frac12\sqrt{\frac{\pi}{a}}\exp\left(\frac{b^2{-}ac}{a}\right)\erf\left(\sqrt{a}x{+}\frac{b}{\sqrt{a}}\right),
\end{align*}
we get \eqref{approx}.

\section{Goodness of Approximation }

\begin{figure}[ht!]
	\centering
	\begin{subfigure}{.49\textwidth}
		\centering
		\!\!\!\includegraphics[width=\linewidth]{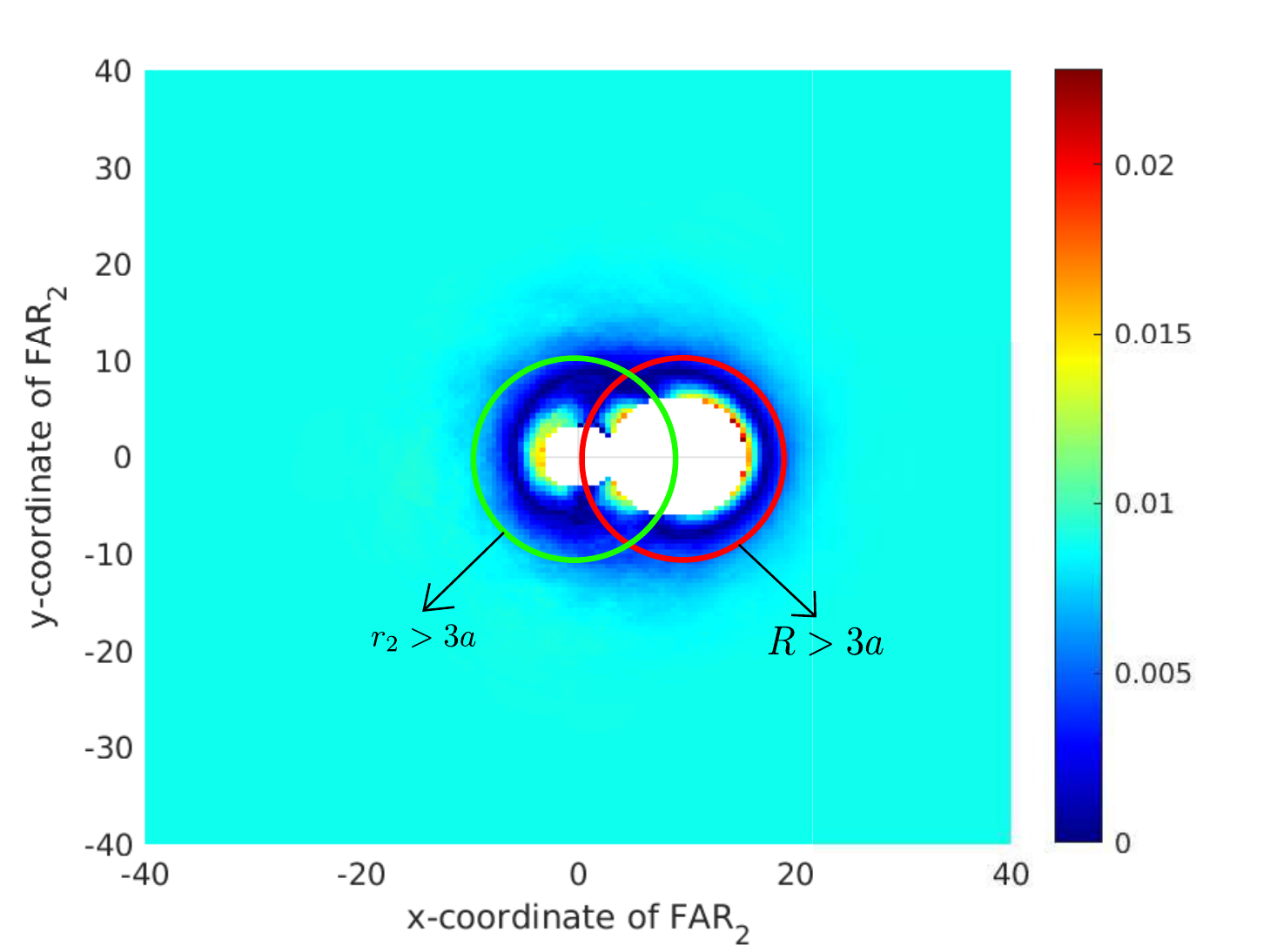}\!\!\!
		\caption{$a=3 \mu $m, $ r_1=9\mu $m}
	\end{subfigure}
	\begin{subfigure}{.49\textwidth}
		\centering
		\!\!\!\!\includegraphics[width=\linewidth]{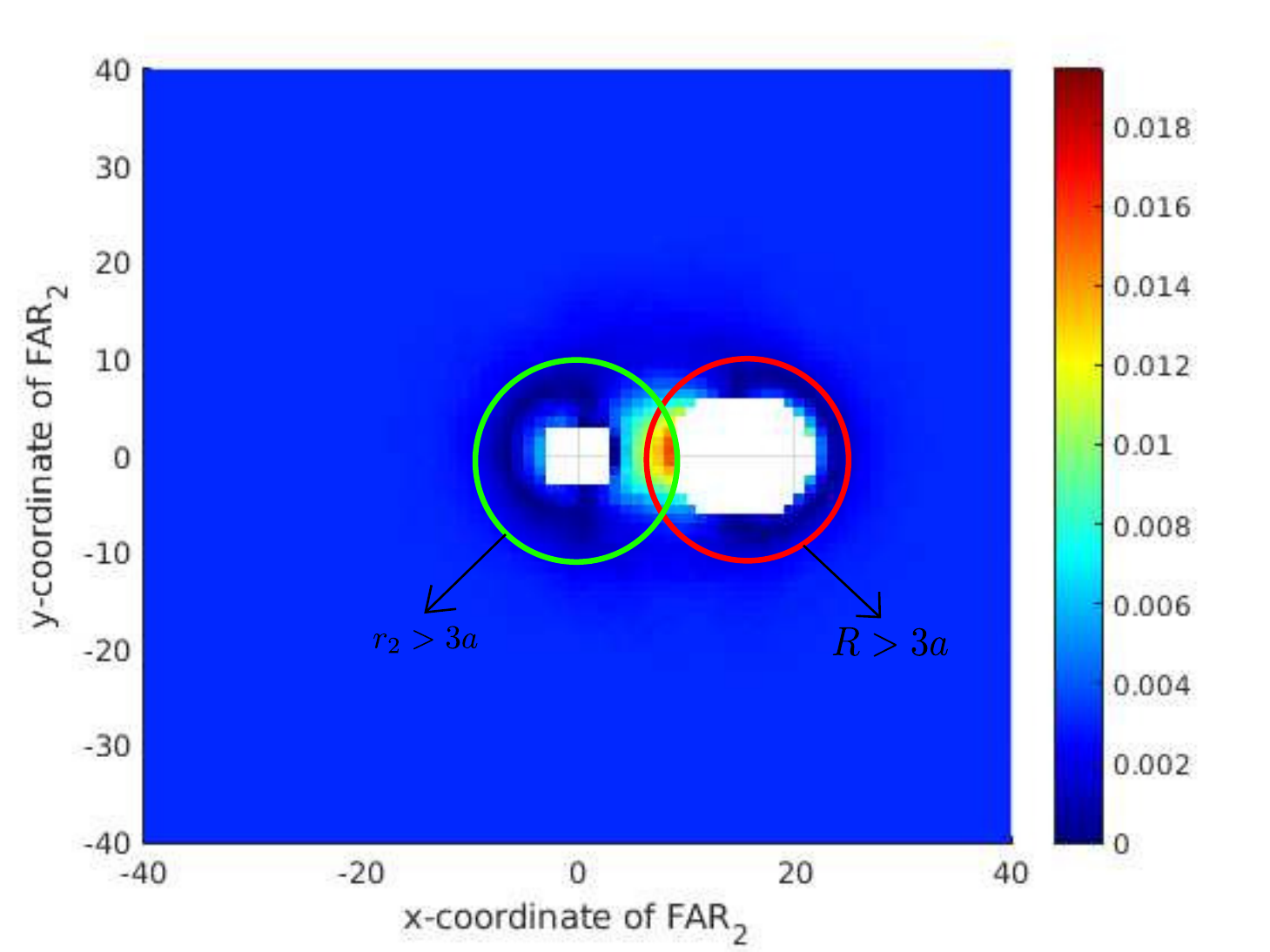}\!\!\!
		\caption{$a=3 \mu $m, $ r_1=15\mu $m}
	\end{subfigure}
	\caption{Approximation error in the  hitting probability of IM on FAR$ _1 $ in the presence of FAR$ _2 $. The error value at a location $(x,y)$ denotes the AE 
		when FAR$_2$ is located at the location $(x,y)$.   Here FAR$ _1 $ is at a fixed position $ \x_1=[r_1\ 0\ 0] $ and FAR$ _2 $ is shifted in the X-Y plane. 
	}
	\label{fig:supp:surf1}
\end{figure}


To understand how accurate the approximation of IM's hitting point at FAR$_j$ by the point $E_j$ in Appendix A is, we  performe extensive simulations by varying locations of FAR$_1$ and FAR$_2$. 
Fig. \ref{fig:supp:surf1} shows the  the absolute error (AE)  in hitting probability of IM on FAR$ _1 $ in the presence of FAR$ _2 $,  defined as
$$\text{AE}= |\text{Analytical value $ - $ Simulation value}|. $$
The error value at a location $(x,y)$ denotes the AE when the FAR$_2$ is located at the location $(x,y)$ while the location of FAR$_1$ is fixed at $[r_1,0,0]$. 
From Fig \ref{fig:supp:surf1}, we can see that AE is small at most places where there is no overlap between the FARs (denoted by the white color). In particular, the AE is negligible when $ r_2>3a $  and $ R>3a $ (denoted by region outside the green and red circles respectively) for $r_1>3a$.
	\bibliographystyle{IEEEtran}
	\bibliography{two_rxr}
\end{document}